\documentclass{basi}
\usepackage{epsfig} 

\begin{document}
\title{Correlation Between The Rise Rate And The Amplitude Of The Solar Magnetic Cycles} 
\author[B. B. Karak \& A. R. Choudhuri]
       {B. B. Karak, \thanks{Department of Physics, Indian Institute of Science, Bangalore 560012, India; e-mail: bidya$\_$karak@physics.iisc.ernet.in} \& A. R. Choudhuri}
\maketitle
\begin{abstract}
We discuss one important aspect of Waldmeier effect which says that the stronger cycles 
rise rapidly than weaker cycles. We studied four different data set of solar activity indices, 
and find strong linear correlation between rise rates and amplitudes of solar activity. 
We study this effect theoretically by introducing suitable stochastic fluctuations in 
our regular solar dynamo model.
\end{abstract}

\begin{keywords}
(Sun:) sunspots; Sun: magnetic fields; Sun: photosphere
\end{keywords}
It is well known that the stronger cycles take less time to rise their peak values 
than the weaker cycles. It is commonly known as Waldmeier effect. However, with ``Waldmeier effect'' 
two different measures are meant by different authors, causing confusion in the literature. The first is
the observation that the rise times of sunspot cycles are 
anti-correlated to their strengths (i.e.\ the stronger cycles have
shorter rise times) (WE1). The second is that the rates of rise of the
cycles are correlated to their strengths (i.e.\ the stronger cycles
rise faster) (WE2). In the next section we present our observational and theoretical results of WE2.
The details of this work can be found in Karak \& Choudhuri (2010, 2011)

We have studied four different data sets: (1) Wolf sunspot numbers
(cycles 12--23), (2) group sunspot numbers (cycles 12--23), (3)
sunspot area data (cycles 12--23) and (4) $10.7$~cm radio flux
(available only for the last 5 cycles).  All data sets have been
smoothed by a Gaussian filter with a FWHM of $1$~yr.
\begin{figure}
\centerline{\resizebox{14cm}{4cm}{\includegraphics{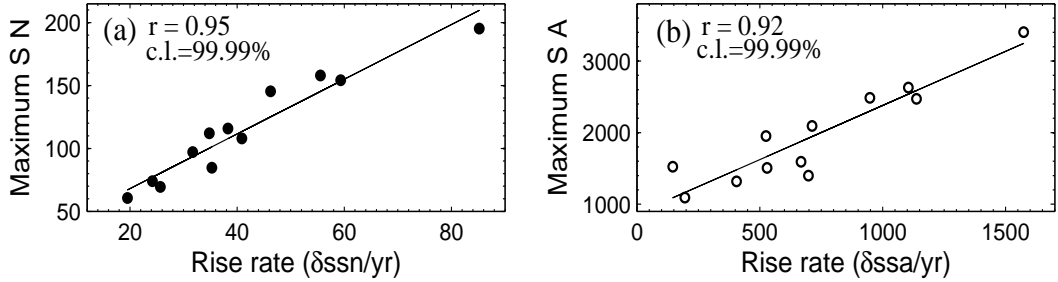}}}
 \caption[]{\label{karak-fig:obser}
(a) Peak sunspot number and (b) sunspot area against rise rates.
}\end{figure}
We calculate the rate of rise by determining the slope between two points at a 
separation of one year, with the first point one year after the sunspot minimum. 
We find strong correlation between the rates of rise and the amplitudes of the sunspot 
cycles. Only the results for sunspot number and sunspot area are shown in panels (a) and (b) of 
Fig.~\ref{karak-fig:obser}.  Cameron \& Schussler (2008) have computed the rise rate slightly 
differently and obtained almost similar results. We conclude that there is a strong 
evidence of WE2 in different kinds of data sets.

\begin{figure}
\centerline{\resizebox{14cm}{4cm}{\includegraphics{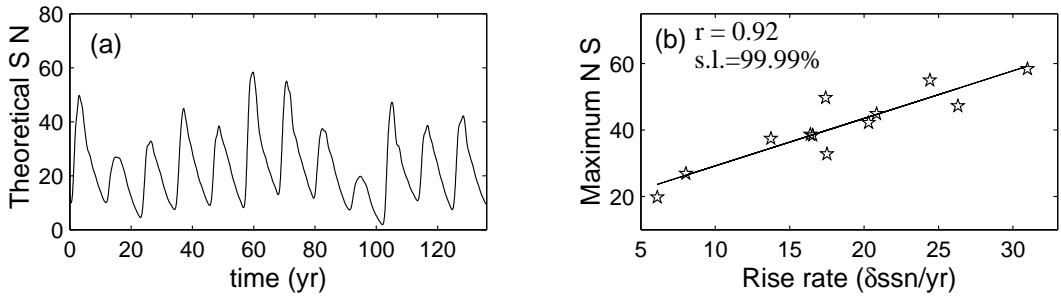}}}
\caption[]{\label{karak-fig:theor}
(a) Theoretical sunspot numbers vs.\ time. (b) Scatter diagram
plotting peak sunspot numbers against rise rates.
}\end{figure}

Now we carry out theoretical study based on our flux transport 
dynamo model (
Chatterjee et al. 2004) to explain this correlation. We believe that the fluctuations in 
the Babcock--Leighton process of poloidal field generation and the fluctuations in the 
meridional circulation are two main sources of irregularities in solar cycles (Jiang et al. 2007; 
Choudhuri \& Karak 2009; Karak 2010). 
Here we show some results in Fig.~\ref{karak-fig:theor} which is obtained by introducing randomness in 
the polar field at each minima in the regular model.


\begin{thebibliography}{10}
\expandafter\ifx\csname natexlab\endcsname\relax\def\natexlab#1{#1}\fi
\bibitem {} Cameron, R., $\&$ Sch\"ussler, M., 2008, 
  \newblock {\it ApJ}, {\bf 685}, 1291
\bibitem {}Choudhuri, A.~R., Karak, B. B., 2011,
  \newblock {\it Res. Astron. Astrophys}, {\bf 9}, 953

\bibitem {} Chatterjee, P., Nandy, D., \& Choudhuri, A.~R., 2004,
  \newblock {\it A$\&$A.}, {\bf 427}, 1019
\bibitem {} Jiang, J., Chatterjee, P., Choudhuri, A.~R., 2007,
  \newblock {\it MNRAS}, {\bf 381}, 1527
\bibitem {} Karak, B. B., 2010,
  \newblock {\it ApJ}, {\bf 724}, 1021
\bibitem {} Karak, B.~B., \& Choudhuri, A.~R., 2010,
  \newblock {in Astrophysics and Space Science Proc., Magnetic
Coupling between the Interior and Atmosphere of the Sun, ed. S. S. Hasan \& R. J. Rutten
(Berlin: Springer)}, {\bf 402}
\bibitem {} Karak, B. B., Choudhuri, A.~R., 2011, 
  \newblock {\it MNRAS}, {\bf 410}, 1503
\end{thebibliography}
\end{document}